\documentclass[showpacs,preprintnumbers,amsmath,amssymb]{revtex4}
\usepackage{graphicx}
\usepackage{dcolumn}
\usepackage{bm}
\usepackage{tensor}
\begin{document}
\title{Helical containers with classical and quantum fluids in rotating frame.}
\author{A.Yu.Okulov} 
\email{alexey.okulov@gmail.com}
\homepage{http://okulov-official.narod.ru}
\affiliation{Russian Academy of Sciences, 
119991, Moscow, Russia}

\date{\ September 30, 2012}

\begin{abstract}
{We consider examples of the classical liquids confined 
by rotating helical boundaries and compare these examples 
with rotating helical reservoir filled by ultracold bosonic ensemble.  
From the point of view of observer who co-rotates with 
classical liquid trapped by 
reservoir the quantum fluid will move translationally 
alongside rotation axis while 
in laboratory frame the quantum fluid will stay in rest. 
This behavior of quantum ensemble 
which is exactly opposite to the classical case might be 
interpreted as a helical analog of Hess-Fairbank effect.}
\end{abstract}


\maketitle
\section{Classical dielectric flow 
at room temperature in helical channel}

Three examples of twisted flows in a rotating reference 
frame are $\mu$ considered $\mu / k_B T $ analytically.
The first case is Stimulated Brillouin scattering in the liquids 
at the room temperature. Here incompressible Navier - Stockes liquid 
\cite {Landau_hydro:1987} is driven in 
rotation by helical interference pattern of the 
counter-propagating optical vortices with opposite angular 
momenta: 
\begin{equation}
\label{Navier_forced}
{\rho \frac {\partial {\vec V}}
{\partial t}}  + \rho
({\vec V} \cdot \nabla){\vec V} =  - \nabla p + { \vec F_{ext}} (\vec r,t){\:}
+\eta \Delta {\vec V}, {\:}{F_{ext}\vec r,t)}= - \nabla \delta p_{str},{\:}
\delta p_{str}= - {\frac {|{\vec E}|^2}{16 \pi}}
\rho_0 {\frac {\partial \epsilon}{\partial \rho }}
\end{equation}
where ${\vec V(\vec r,t)}$ is a field of velocities, 
$\rho (\vec r,t) $ is a liquid density, $p$ is pressure,  $\eta$ is viscosity,
 ${F_{ext}\vec r,t)}= - \nabla \delta p_{str}$ 
is externally applied force which is ponderomotive in this 
particular case \cite {Zeldovich:1985}. The electrostrictive pressure 
$p_{str}$ pulls liquid into regions with a higher optical intensity having 
a form of the $2 \ell $ mutually embedded helices: 
\begin{equation}
\label{TRAPOP1}
\ { |E{(z,r,\theta,t)}|^2}={|E_f|^2}+{|E_b|^2} \sim r^{2|\ell|} 
\exp \bigg{[}-{\frac {2r^2}{D_0^2(1+{ {z^2}/{k^2 D_0^4}})}}\bigg{]}
[1+\cos(2kz+2\ell\theta+\delta \omega t)],{\:}{\:}
\end{equation}
where the cylindrical coordinates $\vec r=(z,r,\theta)$ are used, $D_0$ is 
the radius of LG, $\ell$ is 
vorticity, $k=2 \pi/ \lambda$ is wavenumber, $\delta \omega$ is 
angular Doppler shift 
induced by rotation 
of the reference frame 
or emulated by rotation of Dove prism in phase conjugated 
setup \cite {Okulov:2012josa}.
When their frequencies detuning is adjusted in resonance 
with Brillouin acoustic wave the liquid moves as if it were 
confined inside helical channel having $\lambda/2 = \pi /k $ pitch, 
radius $D_0$ being equal to those of LG beam and the frequency detuning 
being equal to $\delta \omega= 2n V_s \omega_{f,b}/c$, 
where $ V_s $ is the speed of sound, $\omega_{f,b}$ are 
the carrier frequencies of colliding vortices, $c/n$ is 
speed of light in liquid. Typical spatial scales for 
this helical channel are $D_0 \sim 10 - 100 \mu m$, 
$\lambda \sim 0.5 - 2 \mu m$. 
 
The above helical flow at an ambient temperature ${\bf T}= 300 \bf K$ 
appears exactly in the case of the phase-conjugated reflection of the 
optical vortex $ E_f(\vec r,t)$ with orbital angular momentum 
per photon $\ell \hbar$ 
from acoustic wave moving with speed 
$V_s=\sqrt {(\partial p / {\partial \rho})_S}$. In this case a Stockes 
wave $E_b(\vec r,t)$ with downshifted frequency appears 
\cite {Okulov:2008J}. This classical process has a 
quantum counterpart as a decay of a photon with orbital 
angular momentum $\ell \hbar$ to the two particles: 
backward scattered Stockes 
photon with opposite OAM $-\ell \hbar$ and forward 
corkscrew phonon 
with doubled vorticity $ 2 \ell$ \cite {Okulov:2008}. 

For this micrometer-size helical channel width $\lambda/2$ 
and diameter $D_0$ the 
Reynolds number  $\mathcal{R}e=\rho V_s \lambda / \eta$ is 
in the range $10^{(2-3)}$ because of 
high value of velocity $V_s \sim 10^5 cm/sec$ . 
This happens for the viscosities $\eta \sim 10^{-(2-3)} poise$ of 
the most organic solvents and water based 
solutions at the room temperature 
used in applications of the Stimulated Brillouin scattering. 

For $\mathcal{R}e=10^{(2-3)}$ the flow in this channel must be 
indeed turbulent but electrostrictive pressure is strong enough 
to keep acoustic flow inside helical channel formed by a
pair of isolated optical vortices. Moreover even the 
optical speckle field composed of a random set of intertwining 
optical vortices \cite {Okulov:2009} drive the acoustic field 
into rotation exactly at the nodes of the optical interference 
pattern \cite {Okulov:2008}. As a result, the acoustical turbulence 
induced by rotating multiply connected interference pattern 
of the incident optical speckle 
field $E_f(\vec r,t)$ and phase-conjugated replica $E_b(\vec r,t)$ 
is composed of the random set of 
vortex-antivortex pairs \cite {Okulov:2008J}. 

\section{Classical plasma flow in helical channel induced 
by Stimulated Brillouin scattering at $keV$ temperatures}

The laser plasma exhibit strong reflection of compressing radiation 
due to Stimulated Brillouin scattering \cite{Kruer:1990}. We consider 
the ion-acoustic 
wave in an underdense plasma with 
temperature of electrons ${\bf T}_e \sim 10^6 \bf K$
induced again by interference pattern of the 
two counterpropagating optical fields. As in the above described case 
of room temperature dielectric 
we presume that two phase-conjugated optical fields with 
slowly varying envelopes ${{\mathcal {E}}}_{f}$ and ${\mathcal {E}}_{b}$ 
generates ion-acoustic vortex ${\tilde Q} (z,r,\phi,t )$ 
carrying doubled orbital angular momentum due to the motion 
in rotating helical channel. 
The SBS equations are:
\begin{equation}
\label{pumpwave}
\ {\frac {\partial {{{{\mathcal {E}}}_f}(z,r,\phi,t )}} {\partial z} }+
{\frac {n_r} {c} }{\frac {\partial {{{{\mathcal {E}}}_f}}} {\partial t} }+
{\frac {i}{2 k_f}} {\nabla}_{\bot}^{{\:}2} {{{\mathcal {E}}}_f} =
{\frac {i \Omega_p {\:}{n_{e}}} {4{\:} c{\:}{n_c} } } {\tilde Q} {\:}{{\mathcal {E}}}_b 
\end{equation}
\begin{equation}
\label{stockeswave}
\ {\frac {\partial {{{{\mathcal {E}}}_b}(z,r,\phi,t )}} {\partial z} }-
{\frac {n_r} {c} }{\frac {\partial {{{{\mathcal {E}}}_b}}} {\partial t} }-
{\frac {i}{2 k_f}} {\nabla}_{\bot}^{{\:}2} {{{\mathcal {E}}}_b} = -
{\frac {i \Omega_s {\:}{n_{e}}} {4{\:} c{\:}{n_c} } } {{\mathcal {E}}}_f {{\tilde Q}}^{\ast},
\end{equation}
and dimensionless slowly varying ion-acoustic wave complex 
amplitude ${\tilde Q}$ is:
\begin{eqnarray}
\label{acouswave1}
\ {\frac {\partial {{\tilde Q}(z,r,\phi,t )}} {\partial z} }+
{\frac {1} {c_{ia}}} {\frac {\partial {\tilde Q}}  {{\:}\partial t} }
+{\frac {2 {\gamma_{ia}} {{\tilde Q}}} {c_{ia}} }
+{\frac {i}{2 (k_f+k_b)}} {\nabla}_{\bot}^{{\:}2} {\tilde Q}=
&& \nonumber \\
{i(k_f+k_b) } {{{\mathcal {E}}}_f}  
{{{\mathcal {E}}}_b }^{\ast}{\frac {\epsilon_0 }{2{\:} {n_c}{\:}k_B T_{e}}},
\end{eqnarray}
where  ${\nabla}_{\bot}=({\partial_x},{\partial_y})$, 
$n_c \sim 10^(21) cm^{-3} $ is the critical plasma density. 
The resulting plasma flow in 
optically induced corkscrew channel has a form of the phase 
singularity with 
doubled vorticity $2 \ell$ \cite {Okulov_plasma:2010}. The 
electron and ion currents proved to be large enough to produce 
magnetic dipole with kilogauss quasistatic magnetic field $\bf \vec B$. 

\section{Helical channel filled by superfluid at $\mu \bf K$ temperatures}

The macroscopic coherence of quantum fluid in multiply connected geometry 
leads to Hess-Fairbank effect \cite {Fairbank:1967}. The superfluid in 
annular cylindrical container rotating with low angular velocity 
$ \Omega_{\oplus} \sim 10^{-4}  $ rad/sec is not dragged by rotating 
boundaries. This happens when liquid $^4He$ is cooled below 
critical temperature $T_{\lambda}$ 
and angular momentum $<\hat L_Z> \sim N \ell \hbar$ ( rotational 
energy per particle is $<E_{rot}> \sim \ell^2 \hbar^2 / 2 mR^2$) of 
superluid is much smaller than those of classical 
liquid $L_Z \sim N m R^2 \Omega$, where R is a 
mean radius of flow, m is atom mass, N is a number of atoms in 
a rotating ensemble, $\ell$ is a winding number 
\cite {Feynman:1972,Leggett:2001}. 

The interesting analog of the Hess-Fairbank effect may be proposed 
when superfluid is placed in container having corkscrew shape rather than 
cylindrical one \cite {Aldoss:2016}. For this purpose 
not only liquid $^4 He$ is feasible but 
microkelvin trapped alkali gases are suitable as well. 
The Gross-Pitaevskii equation 
is applicable in the latter case. 
The trapping helical optical potential with soft penetrable walls is as follows:
\begin{equation}
\label{TRAPOP1}
\ { U_{opt}{(z,r,\theta,t)}} \sim {\frac 
{U_0 \cdot r^{2|\ell|}}{(1+z^2/(k^2 {D_0}^4))} }
\exp \bigg{[}-{\frac {2r^2}{D_0^2(1+{ {z^2}/{k^2 D_0^4}})}}\bigg{]}
[1+\cos(2kz+2\ell\theta+\delta \omega t)],{\:}{\:}
\end{equation}
where  $\delta \omega$ is 
angular Doppler shift 
induced by rotation 
of the reference frame 
or emulated by rotation of Dove prism in a phase conjugated 
setup \cite {Okulov:2012josa}.
Transformation to the reference frame rotating synchronously 
with angular velocity $\Omega=\delta \omega /2 \ell=|\vec \Omega_{\oplus}|$ 
with trapping helix leads to the time-dependent Gross-Pitaevskii equation 
(GPE) \cite {Pitaevskii:1999,Okulov_helical:2012,Berloff:2008}: 
\begin{equation}
\label{GPE_rot_frame}
{i \hbar}{\frac {\partial {\Psi}}
{\partial t}} = -{\frac {\hbar^2}{2 m}} 
\Delta {\Psi} +{ \tilde U_{opt}{(z,r,\theta)}}{\:}{\Psi} 
+{g}
|{\Psi}|{\:}^2 {\Psi}-\Omega \hat L_z {\Psi},
\end{equation}
where the stationary wavefunctions for the superfluid 
ensemble $\Psi=\Phi(z,r,\theta)\exp(-i\mu t/ \hbar)$ 
are given by: 
\begin{equation}
\label{GPE_static_in_rot_frame}
\mu {\Phi} = -{\frac {\hbar^2}{2 m}} 
\Delta {\Phi} +{\tilde  U_{opt}{(z,r,\theta)}}{\:}{\Phi} 
+{g}|{\Phi}|{\:}^2 {\Phi}+
\Omega {\hbar}{\frac {\partial {\Phi}}{\partial \theta}},{\:}
g={\frac {4 \pi \hbar^2 a_S}{m}}, 
\end{equation}
where $a_S$ is $S$-wave scattering length. 
We evaluate linear and angular momenta of the superfluid 
ensemble in a helical 
container \cite {Okulov_helical:2012} and discuss 
the possibilities of rotations detection with this geometry. 
Noteworthy the "observer" velocity 
vector $\vec V$ with respect to "lab frame" 
has two components \cite {Okulov_plasma:2010}: the 
azymuthal velocity 
$\vec V_{\theta}= \vec {\Omega_{\oplus}} \times \vec r$ stands for 
helix rotation around LG propagation axis $\vec z$ while helix 
pitch velocity $\vec V_{z}= ({\vec z}/z) {\delta \omega}/ 2 k$ 
is responsible for wavetrain translation 
along $\vec z$ \cite {Okulov:2013}.  

\section{Discussion}

It is shown in the first two examples that certain classical 
liquids in helical container 
are completely dragged by rotating boundaries, so that observer 
placed in the reference frame collocated with rotating container 
will not detect rotation. On the contrary the quantum fluid 
placed in slowly rotating container will 
remain in rest in laboratory frame 
and it will move translationally  from the point of view of 
observer placed in rotating frame collocated with dragged 
classical fiud. Experimentally the micrometer size 
corkscrew channels may be realized as interference patterns 
of detuned optical vortices (for a trapped degenerate quantum gas) 
or as a twisted glass 
pipe (for a $^4 He$ cooled below $\lambda$-point)
\cite {Kapitza:1938,Allen:1938,Kapitza:1941}.


\begin{thebibliography}{99}

\bibitem{Landau_hydro:1987} 
{L.D. Landau and E.M. Lifshitz},\textit 
{"Fluid Mechanics"},{ Butterworth-Heinemann, Oxford}(1987).

\bibitem{Zeldovich:1985}
{B.Y.Zeldovich, N.F.Pilipetsky and V.V.Shkunov} 
{\itshape "Principles of Phase Conjugation"},Ch.2,  
(Berlin:Springer-Verlag )(1985). 

\bibitem{Okulov:2012josa}{A.Yu.Okulov,} J. Opt. Soc. Am. B 
{\bf 29}, 714-718 (2012).

\bibitem{Okulov:2008J}{A.Yu.Okulov,} JETP Lett., {\bf 88}, 631 (2008).
 
\bibitem{Okulov:2008}
{A.Yu.Okulov,} J.Phys.B., {\bf 41},101001 (2008).

\bibitem{Okulov:2009} {A.Yu.Okulov,} {Phys.Rev.A }, 
{\bf 80}, 013837 (2009).

\bibitem{Kruer:1990}W.L. Kruer , \textit 
{"The Physics of Laser Plasma Interactions"}, 
Addison-Wesley, New York (1990).
  
\bibitem{Okulov_plasma:2010}{A.Yu.Okulov,} 
Phys.Lett.A, {\bf 374},4523-4527 (2010).

\bibitem{Fairbank:1967} 
{G. Hess and W. Fairbank,}{Phys.Rev.Lett.}, {\bf 19}, 216 (1967).

\bibitem{Feynman:1972}
{R.P.Feynman,}{\itshape "Statistical mechanics"}, 
Ch.11,(1972) Reading, Massachusetts.

\bibitem{Aldoss:2016} {Anwar Al Rsheed, 
Andreas Lyras, Omar M. Aldossary, and Vassilis E. Lembessis} 
{Phys.Rev.A }, {\bf 94}, 063423 (2016). 

\bibitem{Leggett:2001} 
{A.J. Leggett}, {Rev.Mod.Phys.} {\bf 73}, 307-356 (2001).
 
\bibitem{Pitaevskii:1999}
{F. Dalfovo, S.Giorgini, S.Stringari, L.P.Pitaevskii},
{Rev.Mod.Phys.}{\bf 71},463(1999).

\bibitem{Okulov_helical:2012} {A.Yu.Okulov,} {Phys.Lett.A}, {\bf 376}, 
650-655 (2012).

\bibitem{Berloff:2008} {J.Keeling and N. G. Berloff},
{Phys.Rev.Lett.,} {\bf 100},250401 (2008).

\bibitem{Okulov:2013} {A.Yu.Okulov,}
"Superfluid rotation sensor with helical laser trap",  
{Journ.Low.Temp.Phys.}, {\bf 171}, 397-407 (2013).

\bibitem{Kapitza:1938}{P.L.Kapitza,} Nature, {\bf 141}, 74 (1938).

\bibitem{Allen:1938}{J.F.Allen, A.D.Misener,} Nature, {\bf 141}, 74 (1938).

\bibitem{Kapitza:1941} {P.L.Kapitza,} Phys.Rev., {\bf 60}, 354 (1941).

\end{thebibliography}
\end{document}